\documentclass[10pt]{article}

\usepackage{graphicx}

\begin{document}
\title{\centerline{\textbf{\large{Lagrangian drifter dispersion in the southwestern Atlantic Ocean}}}}

\author{
\centerline{\textsc{Stefano Berti}}\\
\centerline{\textit{\footnotesize{Laboratoire Interdisciplinaire de Physique, Grenoble, and 
Laboratoire de M\'et\'eorologie Dynamique, Paris, France}}}
\and 
\centerline{\textsc{Francisco Alves Dos Santos}}\\
\centerline{\textit{\footnotesize{PROOCEANO Servi\c{c}o Oceanograf\'ico, Rio de Janeiro, Brazil}}}
\and
\centerline{\textsc{Guglielmo Lacorata}}\\
\centerline{\textit{\footnotesize{Institute of Atmospheric and Climate Sciences, National Research Council,
Lecce, Italy}}}
\and 
\centerline{\textsc{Angelo Vulpiani}}\\
\centerline{\textit{\footnotesize{Department of Physics, CNR-ISC and INFN, Sapienza University of Rome, Rome, Italy}}}
}

\date{}
\maketitle

\begin{abstract}
In the framework of Monitoring by Ocean Drifters (MONDO) Project, a set of Lagrangian drifters were released in proximity of the Brazil Current, the western branch of the Subtropical Gyre in the South Atlantic Ocean. The experimental strategy of deploying part of the buoys in clusters offers the opportunity to examine relative dispersion on a wide range of scales.   
Adopting a dynamical systems approach, we focus our attention on scale-dependent indicators, like the finite-scale Lyapunov exponent (FSLE) and the finite-scale (mean square) relative velocity (FSRV) between two drifters as function of their separation, and compare them with classic time-dependent statistical quantities like the mean square relative displacement between two drifters and the effective diffusivity as functions of the time lag from the release.  
We find that, dependently on the given observable, the quasigeostrophic turbulence scenario is overall compatible with our data analysis, with discrepancies from the expected behavior of 2D turbulent trajectories likely to be ascribed to the non stationary and non homogeneous characteristics of the flow, as well as to possible ageostrophic effects.  
Submesoscale features of $\sim O(1)$ km are considered to play a role, to some extent, in determining the properties of relative dispersion as well as the shape of the energy spectrum.  We present, also, numerical simulations of an OGCM of the South Atlantic, and discuss the comparison between experimental and model data about mesoscale dispersion. 
\end{abstract}

\section{Introduction}

Detailed investigation of Geophysical flows involves experimental campaigns in which buoys, in the ocean, or balloons, in the atmosphere, are released in order to collect Lagrangian data against which theories and models can be tested.  Questions concerning oil spill fate, fish larvae distribution or search and rescue operations are only a few examples that make the study of advection and diffusion properties not only a challenging scientific task, but also a matter of general interest. 

\begin{figure}[!t]
\begin{center}
\includegraphics[clip=true,width=0.65\textwidth,angle=0]{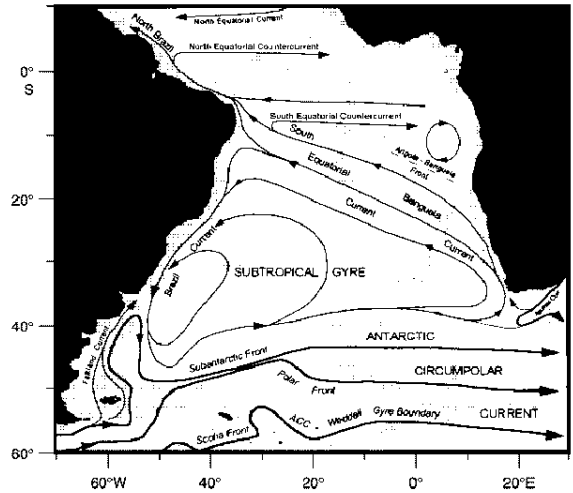}
\end{center}
\caption{Large-scale surface circulation in SAO (courtesy of Integrated Ocean Drilling Program; available online at http://www.iodp.org/).}
\label{fig:sao}
\end{figure}

In the past years, an amount of Lagrangian data about the South Atlantic Ocean (SAO)  was collected thanks to the First Global Atmospheric Research Program (GARP) Global Experiment (FGGE) drifters,  released following the major shipping lines, the Southern Ocean Studies (SOS) drifters, deployed in the Brazil-Malvinas Confluence (BMC) and the Programa Nacional de B\'oias (PNBOIA) drifters [Brazilian contribution to the Global Oceans Observing System (GOOS)], released in the Southeastern Brazilian Bight (SBB). These data allowed estimates of Eddy Kinetic Energy (EKE), integral time scales and diffusivities  (Piola et al. 1987; Figueroa and Olson 1989; Sch\"afer and Krauss 1995). Despite the relatively uniform coverage, the boundary currents resulted poorly populated by buoys; furthermore, all previous studies about drifters in the South Atlantic have concerned  one-particle statistics only.   
In this regard, in the framework of Monitoring by Ocean Drifters (MONDO) Project, a recent Lagrangian experiment, consisting in the release of a set of 39 World Ocean Circulation Experiment (WOCE) Surface Velocity Program (SVP) drifters, was planned in relationship with an oil drilling operation in proximity of the coast of Brazil, around ($24^{\circ}$S, $44^{\circ}$W). Part of the drifters were deployed in 5-element clusters, some of them with initial drifter separations smaller than 1 km.   
This set of satellite-tracked Lagrangian trajectories offers, now, the opportunity to revisit advective and diffusive properties characterizing the current systems explored by the drifters. From the analysis of trajectory pair dispersion we can extract, in principle, information about the dominant physical mechanism acting at a certain scale of motion (e.g. chaotic advection, turbulence, diffusion). A thorough description of the oceanography of the South Atlantic Ocean, particularly of the main circulation patterns and of the mass transport properties, can be found in  Peterson and Stramma (1991); Campos et al. (1995); Stramma and England (1999). The major feature characterizing the central region of the SAO  is the large anticyclonic (anticlockwise) circulation known as Subtropical Gyre (SG). Other relevant surface current systems are: South Equatorial Current (SEC), Brazil Current (BC), Malvinas Current (MC), South Atlantic Current (SAC) and Benguela Current (BgC), as shown in Fig. \ref{fig:sao}. 

\begin{figure}[!t]
\begin{center}
\includegraphics[clip=true,width=0.7\textwidth,angle=0]{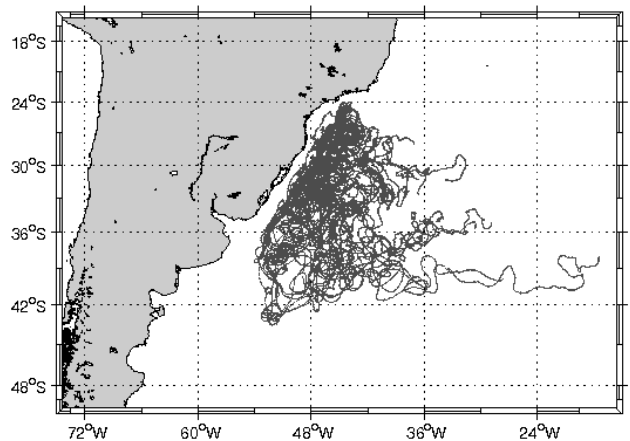}
\end{center}
\caption{Overall view of the trajectories of MONDO drifters.}
\label{fig:drifters}
\end{figure}

In the quasigeostrophic (QG) approximation, valid for relative vorticities much smaller than the ambient vorticity because of the earth's rotation, some theoretical arguments would predict that, from the scale of the forcing at which eddies are efficiently generated by instability, e.g. the Rossby radius of deformation,  both a  down-scale enstrophy cascade and an up-scale energy cascade take place, corresponding to energy  spectra $E(k) \sim k^{-3}$ and $E(k) \sim k^{-5/3}$, respectively (Kraichnan 1967; Charney 1971). 
From a relative dispersion perspective, in the forward cascade range, the mean square relative displacement between two trajectories grows exponentially fast in time 
(Lin 1972), whereas, in the inverse cascade range, it follows the $\sim t^3$ power law (Obhukov 1941; Batchelor 1950). Possible deviations from this ideal picture may reasonably come from the non homogeneous and non stationary characteristics of the velocity field: for example, in the case of boundary currents, as well as from ageostrophic effects.   At this regard, one presently debated issue is the role of submesoscale vortices (McWilliams 1985) [velocity field features of size $\sim O(1)$ km] in determining the shape of the energy spectrum at intermediate  scales between the Rossby deformation radius [in the ocean typically $\sim (10-50)$ km] and the dissipative scales (much smaller than 1 km). A thorough discussion about submesoscale processes and dynamics can be found in Thomas et al. (2008). 
Recent high-resolution 3D simulations of upper layer turbulence  (Capet et al. 2008a,b;  Klein et al. 2008) have shown that the direct cascade energy spectrum flattens from $k^{-3}$  to  $k^{-2}$  for order $O(1)$ Rossby number $R_o=U/fL$, where $U$ is the typical velocity difference on a characteristic length $L$ of the flow and $f \simeq 10^{-4}$ $s^{-1}$ is the Coriolis parameter. 

Our main purpose is to exploit the MONDO drifter trajectories, shown in Fig.~\ref{fig:drifters}, to examine relative dispersion by means of several indicators,  and discuss the consistency of our data analysis in comparison with classical turbulence theory predictions, model simulations and previous drifter studies available for different regions of the ocean.  
   
This paper is organized as follows: in section \ref{sec:diffusion} we recall the definitions of the major indicators of the Lagrangian dispersion process; in section \ref{sec:data} we give a description of the MONDO drifter Lagrangian data; in section  \ref{sec:results} the outcome of the data analysis is presented;  in section \ref{sec:model}, we report the analysis of the ocean model Lagrangian simulations in comparison with the observative data; and, in section \ref{sec:conclusions}, we outline and discuss the main results we have obtained in the present work.    
\section{Lagrangian dispersion indicators}
\label{sec:diffusion}

\subsection{One-particle statistics}
\label{sec:absolute}
 
Let $\mathbf{r}=(x,y,z)$ be the position vector of a Lagrangian particle, in a 3D space, evolving according to the equation $d\mathbf{r}/dt = \mathbf{U}(x,y,z,t)$, 
where $\mathbf{U}(x,y,z,t)$ is a 3D Eulerian velocity field, and let us indicate with $\mathbf{v}(t)$ the Lagrangian velocity along the trajectory $\mathbf{r}(t)$.   
Let us imagine, then, a large ensemble of $N \gg 1$ Lagrangian particles, passively advected by the given velocity field, and refer, for every statistically averaged quantity, to the mean over the ensemble.   

The autocorrelation function of a Lagrangian velocity component $v$ can be defined, for $t_1 \geq t_0$, as:
 \begin{equation}
  C(t_1,t_0) =\left[ \langle   v(t_1)v(t_0) \rangle - \langle v(t_0) \rangle^2 \right] / \left[ \langle v(t_0)^2 \rangle - \langle v(t_0) \rangle^2 \right]
  \label{eq:autocorr}
\end{equation}
In case of stationary statistics, $C(t_1,t_0)$ depends only on the time lag $t=t_1-t_0$. The integral Lagrangian time $\tau_L$ is the time scale after which 
the autocorrelation has nearly relaxed to zero. Typically $\tau_L$ can be estimated as the time of the first zero crossing of $C(t_1,t_0)$, or, alternatively, as 
the time after which $|C(t_1,t_0)|$ remains smaller than a given threshold.  

Absolute dispersion can be defined as the variance of the particle displacement relatively to the mean position at time $t$:
 \begin{equation}
  \langle A^2(t) \rangle = 
   \langle \left[ \mathbf{r}(t)-\mathbf{r}(0) \right]^2 \rangle - \langle \left[ \mathbf{r}(t)-\mathbf{r}(0) \right]  \rangle^2 
  \label{eq:absdisp}
\end{equation}
In the limit of very small times, absolute dispersion is expected to behave as follows:
\begin{equation}
  \langle A(t)^2\rangle
  \simeq \sigma^2_L t^2, \,\,\,\, \mathrm{for} \,\,t \ll \tau_L
  \label{eq:ballistic}
\end{equation}
where $\tau_L$ is the Lagrangian autocorrelation time and $\sigma^2_L$ is the total Lagrangian velocity variance. 
The ballistic regime (\ref{eq:ballistic}) lasts as long as the trajectories save some memory of their initial conditions. 
In the opposite limit of very large times, when the autocorrelations have relaxed to zero and the memory of the initial conditions is lost, we have:
\begin{equation}
  \langle A(t)^2 \rangle
 \simeq 2 K  t, \, \, \, \, \mathrm{for} \, \, t \gg \tau_L
  \label{eq:brown}
\end{equation}
where $K$ is the absolute diffusion coefficient~(Taylor 1921). 

Although single particle statistics give information about the advective transport, mostly because of the
largest and most energetic scales of motion, two (or more) particle statistics give information about
the physical mechanisms acting at any scale of motion, compatibly with the available resolution.
\subsection{Two-particle statistics}
\label{sec:relative}

Let us indicate with $R(t)=||\mathbf{r}^{(1)}(t)-\mathbf{r}^{(2)}(t)||$ the distance between two trajectories at time $t$. 
Relative dispersion is defined as the second order moment of $R(t)$:
\begin{equation} 
  \langle R^2(t) \rangle =
  \langle ||\mathbf{r}^{(1)}(t)-\mathbf{r}^{(2)}(t)||^2 \rangle,
  \label{eq:reldisp}
\end{equation}
where the average is over all the available trajectory pairs $(\mathbf{r}^{(1)}, \mathbf{r}^{(2)})$. 

In the small scale range, the velocity field between two sufficiently close trajectories is reasonably assumed to vary smoothly.   
This means that, in nonlinear flows, the particle pair separation typically evolves following an exponential law: 
\begin{equation}
  \langle R^2(t) \rangle \sim e^{L(2) t} 
  \label{eq:chaos}
\end{equation}
where, from the theory of dynamical systems, $L(2)$ is the Generalized Lyapunov Exponent of order 2 (Bohr et al. 1998). 
When fluctuations of the finite-time exponential growth rate around its mean value are weak, one has 
$L(2) \simeq 2 \lambda_L$, where $\lambda_L$ is the (Lagrangian) Maximum Lyapunov exponent (MLE; Boffetta et al. 2000).  
Notice that for ergodic trajectory evolutions the Lyapunov exponents do not depend on the initial conditions. 
If $\lambda_L > 0$ (except for a set of zero probability measure) we speak of Lagrangian chaos. 
The chaotic regime holds as long as the trajectory separation remains sufficiently 
smaller than the characteristic scales of motion.  

In the opposite limit of large particle separations, when two trajectories are sufficiently distant
from each other to be considered uncorrelated, the mean square relative displacement behaves as: 
\begin{equation}
\langle R^2(t) \rangle \simeq 4 K_E t, \, \, \, \, \mathrm{for} \, \, t \to \infty 
  \label{eq:reldiff}
\end{equation}
where we indicate with $K_E$ the asymptotic eddy-diffusion coefficient (Richardson 1926). 

At any time $t$, the diffusivity $K(t)$ can be defined as:
\begin{equation}
    K(t) = \frac{1}{4} \langle \frac{dR^2}{dt}(t) \rangle =  \frac{1}{2} \langle R(t) \frac{dR}{dt}(t) \rangle
  \label{eq:diffusivity}
\end{equation}  
with $K(t) \to K_E$ for $t \to \infty$. 

If the velocity field is characterized by several scales of motion, relative dispersion in the intermediate 
range, i.e. between the smallest and the largest characteristic length, depends on the properties of 
local velocity differences, i.e. the mean gradients on finite scale. For instance, 
in 3D fully developed turbulence (Frisch 1995), relative dispersion follows the so-called 
Richardson's law: 
 \begin{equation}
  \langle R^2(t) \rangle \sim  t^{\gamma}, 
  \label{eq:richardson}
\end{equation}
with $\gamma=3$, as long as the trajectory separation lies in the inertial range of the energy 
cascade~(Richardson 1926) from large to small scales. 
It is worth to remark that Richardson's law also holds in the inverse cascade range (from small to large scales) of 
2D turbulence because, in that case as well, the energy spectrum follows Kolmogorov's $k^{-5/3}$ scaling, exactly as in the 
inertial range of 3D turbulence (Kraichnan 1967).
Any power law of the type (\ref{eq:richardson}) for $\langle R^2(t) \rangle$ with $\gamma > 1$ is 
known as super-diffusion. 
\subsection{Scale-dependent indicators}
\label{sec:FSLE}

The Finite-Scale Lyapunov Exponent (FSLE) has been formerly introduced as 
the generalization of the MLE $\lambda$ for non-infinitesimal perturbations ~(Aurell et al. 1996). 
If $\delta$ is the size of the perturbation on a trajectory in the phase space of a system, 
and $\langle \tau(\delta) \rangle$ is the phase space averaged time
that $\delta$ takes to be amplified by a factor $\rho > 1$, then the FSLE is defined as
\begin{equation}
  \lambda(\delta) = {1 \over \langle \tau(\delta) \rangle} \ln \rho
  \label{eq:fsle}
\end{equation}
The quantity  $\tau(\delta)$ is the exit time of the perturbation size from the scale $\delta$, and it is defined as 
the first arrival time to the scale $\rho \cdot \delta$, with $\rho \sim O(1)$. 
The computation of the expectation value 
of the growth rate at a fixed scale, which justifies the definition (\ref{eq:fsle}), is described 
in~Boffetta et al. (2000).  
As far as Lagrangian dynamics are concerned, the evolution equations of the Lagrangian trajectories form a 
dynamical system whose phase space is the physical space spanned by the trajectories. In this context, 
the analysis of relative dispersion can be treated as a problem of finite-size perturbation evolution, with 
scale-dependent growth rate measured by the FSLE. 
The first who had the idea to measure the relative dispersion, or, equivalently, 
the diffusivity, as a function of the trajectory separation was Richardson (1926). The FSLE is fundamentally based on the same principle. 
Recently, the use of fixed-time and fixed-scale averaged indicators of relative dispersion in various contexts, 
from dynamical systems to observative data in ocean and atmosphere, have been reviewed  
and discussed in several works (Artale et al. 1997; Boffetta et al. 2000; Lacorata et al. 2001, 2004; LaCasce and Ohlmann 2003, 
LaCasce 2008).  
By a dimensional argument, if relative dispersion follows a $\langle R^2(t) \rangle \sim
t^{2/\beta}$ scaling law, then the FSLE is expected to scale as $\lambda(\delta) \sim
\delta^{-\beta}$. For example, in the case of standard diffusion we expect 
$\beta=2$; in Richardson's super-diffusion, $\beta=2/3$; in ballistic or shear dispersion we have $\beta=1$. 
Chaotic advection means exponential separation between trajectories. In terms of FSLE this amounts to a scale-independent $\lambda(\delta)=$constant: that is, $\beta \to 0$. 
In the limit of infinitesimal separation, the FSLE is nothing but the MLE, i.e. $\lambda(\delta) \simeq \lambda_L$ (Aurell et al. 1996). Under these conditions, relative dispersion is said to be a non-local process, since it is determined by velocity field structures with a characteristic scale much larger than the particle separation. On the contrary, when the growth of the distance between two particles is mainly driven 
by velocity field structures of the same scale as the particle separation, relative dispersion is said to be a local process. 
The super-diffusive processes occurring in 2D and 3D turbulence are phenomena of this type. 

An indicator related to the FSLE is the mean square velocity difference between two trajectories as function of their separation. 
Indicating with ${\mathbf r}^{(1)}$, ${\mathbf r}^{(2)}$, ${\mathbf v}^{(1)}$, ${\mathbf v}^{(2)}$ the positions and the Lagrangian 
velocities, respectively, of two particles $1$ and $2$ at a given time,  we define the Finite-Scale Relative Velocity (FSRV) at scale $\delta$:
\begin{equation}
  \langle \left[ \Delta V(\delta) \right]^2 \rangle = \langle  \left[  {\mathbf v}^{(1)} -  {\mathbf v}^{(2)} \right]^2  \rangle 
  \label{eq:fsrv}
\end{equation}
where the average is over all trajectory pairs fulfilling the condition $R(t)=||{\mathbf r}^{(1)}(t) - {\mathbf r}^{(2)}(t)||=\delta$ at some time $t$. 
From the FSRV a scale-dependent diffusivity can be formed as $K(\delta)  = (1/2) \cdot \delta \cdot \langle \left[ \Delta V(\delta) \right]^2 \rangle^{1/2}$ and compared to the classical time-dependent diffusivity $K(t)$ defined in (\ref{eq:diffusivity}).      

We would like to end this section with a remark. It is physically reasonable to assume that, for statistically homogeneous flows, the behavior of the relative dispersion changes in correspondence of certain characteristic lengths of the system, rather than  after certain time intervals.  The major shortcoming when measuring the mean square relative displacement at fixed time consists in averaging together particle pairs that may have very different separations, 
i.e. may belong to very different dispersion regimes. 
On the contrary, the mean dispersion rate at fixed scale, or other scale-dependent related quantities, generally have the 
benefit of being weakly affected by possible overlap effects.  This is of particular relevance since the presence of scaling laws 
in the pre-asymptotic regime raises questions about the characteristics (e.g., the kinetic energy spectrum) of the underlying 
velocity field.  

\section{Data set and local oceanography}
\label{sec:data}

The MONDO Project is a campaign planned by PROOCEANO for ENI Oil do Brasil to provide unrestricted access to oceanographic data and subsidize not only the companies operational goals but also the scientific community.  The Lagrangian experiment consisted in the deployment of 39 WOCE SVP drifters (Sybrandy and Niiler 1991) in the surroundings of an oil drilling operation ($24^{\circ}$ 24' 50" S and $43^{\circ}$ 46' 22" W) in proximity of the BC. 
The drifters are of the holey-sock type, composed by a surface floater, a tether and a submerged drogue, following the design criteria proposed by Sybrandy and Niiler (1991). The drogue length is 6.44 m, centered at 15 m deep, in order to represent the first 20 m current information. 
The drifters are equipped with a GPS device and IRIDIUM transmission, providing a positioning accuracy of 7 meters and an acquiring rate of 3 hours. From the 39 drifters, 14 were deployed individually in a 3-day frequency and 25 were deployed in clusters of 5, every 12 days. Justification for the 3-day frequency deployment is based upon the 6.5 days energy peak in the wind variability in the region 
(Stech and Lorenzetti 1992). Data used in this study comprehend the period within 21 September 2007 - 14 November 2008 and passed the quality control proposed by Hansen and Poulain (1996) in the Global Drifter Program (GDP) database. High frequency components have been removed by applying a Blackman low-pass filter with a window of 15 points (45 hours). 

The domain explored by the MONDO Project drifters mainly corresponds to that of the BC and to the area 
where the southward flowing BC meets the northward flowing MC, forming the BMC, 
which results in an eastward current feeding the SAC. BC is a warm western boundary current, flowing southward and 
meandering over the 200 m isobath (Peterson and Stramma 1991; Lima et al. 1996). As reported by de Castro (2006) and Garfield (1990) 
the velocities inside the BC have values between 25 and 80 $\mathrm{cm\;} \mathrm{s}^{-1}$. BC flux increases with 
latitude (Assireu 2003; M\"uller et al. 1998; Gordon and Greengrove 1986) and 
veers towards east when meeting the colder Malvinas Current around 40$^{\circ}$S. The BMC is a highly energetic zone with strong 
thermal gradient and great variability, playing an important role in weather and climate of 
South America (Legeckis and Gordon 1982; Pezzi et al. 2005; Piola et al. 2008). 
Both BC and BMC show intense mesoscale activity with eddies detaching from both sides of the flow. Lentini et al. (2002) used 
satellite derived Sea Surface Temperature (SST) to estimate an average of 7 rings per year, with lifetimes ranging from 11 
to 95 days with major and minor radii of 126 $\pm$ 50 km and 65 $\pm$ 22 km respectively. Assireu (2003) estimates 
a diffusion coefficient between $6 \cdot 10^6$ and $9.1 \cdot 10^7$ $\mathrm{cm}^2\mathrm{\;s}^{-1}$, 
a Lagrangian time scale between 1 and 5 days and a characteristic (Eulerian) length scale varying from 19 to 42 km, which agrees with the meridional 
variation of the 1$^\mathrm{st}$ internal Rossby radius of deformation in the region (Houry et al. 1987).
Studies addressing the energetics in BC and BMC flows report great spatial and temporal variability 
(see, e.g., Assireu et al. 2003; Piola et al. 1987; Stevenson 1996) with EKE varying from 200 to 
1200 $\mathrm{cm}^2\mathrm{\;s}^{-2}$ and representing 70-90\% of the Total Kinetic Energy (TKE). Recently, Oliveira et al. (2009) 
extended the analysis made by Piola et al. (1987) and suggested that EKE is usually lower than Mean Kinetic Energy (MKE) in the BC and always greater in the BMC.

\section{Data analysis}
\label{sec:results}

\subsection{Single particle statistics}
\label{sec:single}

\begin{figure}[!t]
\begin{center}
\includegraphics[clip=true,width=0.7\columnwidth,angle=0]{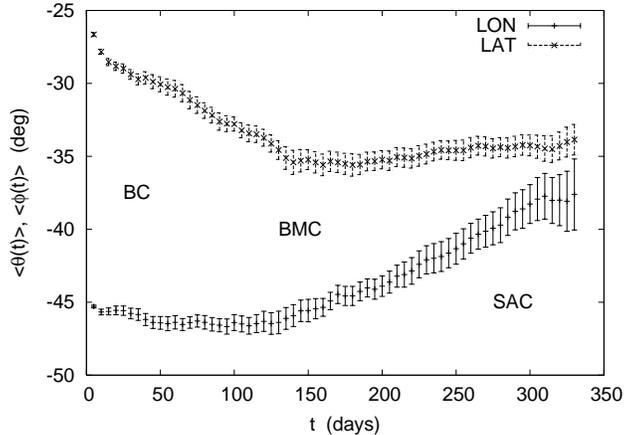}
\end{center}
\caption{Mass center coordinates in longitude $\langle \theta \rangle$ and latitude $\langle \phi \rangle$, averaged over all drifters, 
as function of the time lag since the release. The time sampling is $\Delta t=1/8$~day. Error bars are the standard deviations on the mean values.}
\label{fig:lonlat}
\end{figure}
\begin{figure}[!t]
\begin{center}
\includegraphics[clip=true,width=0.7\textwidth,angle=0]{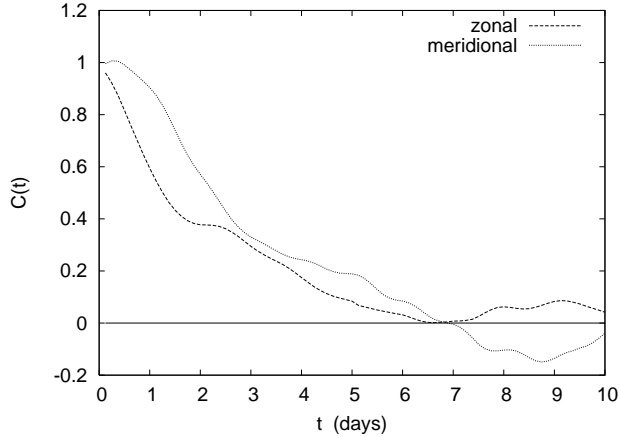}
\end{center}
\caption{Lagrangian velocity autocorrelation function for zonal and meridional components. The time sampling is $\Delta t=1/8$~day. 
}
\label{fig:autocorr}
\end{figure}

From a pre-analysis we eliminated two drifters that very soon lost their drogue, so the set of active drifters is reduced to 37 units.   
The Lagrangian velocity statistics, reconstructed along the drifter trajectories by means of a finite difference scheme,  
 provides standard deviations, in the zonal and meridional directions, of order $\simeq 25$~cm~s$^{-1}$, compatible with estimates obtained 
from other observative data available for the same ocean region (Figueroa and Olson 1989; Sch\"afer and Krauss 1995). 
     
The drifters initially tend to move southward, driven by the BC; then 
they are nearly stopped by the BMC, and, last, they tend to be transported eastward by the SAC.  
The time evolution of the drifter mean position coordinates in longitude and latitude is shown in Fig.~\ref{fig:lonlat}.  
During the first phase of dispersion, the mean meridional displacement grows almost linearly southward (BC), while 
the mean zonal displacement is almost constant;  
after about 120 days from the release time, the mean meridional displacement tends to saturate (BMC), while the mean 
zonal displacement tends to grow almost linearly eastward (SAC). 
Given the differences between BC, BMC and SAC in terms of energy and circulation patterns, and, in general, between the boundary of the Subtropical Gyre and its interior, the drifters do not sample a statistically homogeneous and stationary domain.  
In order to evaluate the Lagrangian autocorrelations, we divide the trajectory lifetime into time windows of $10$ days width. Then we compute ($\ref{eq:autocorr}$) where mean and variance of the Lagrangian velocity components are recalculated for each window, and then we take the average, for a fixed time lag $t=t_1-t_0$, over all the windows for all trajectories. The resulting functions are plotted in Fig. \ref{fig:autocorr}. 
We notice that, since most of the statistics regards trajectory segments within or near the BC, the motion is slightly more autocorrelated in the meridional than in the zonal direction. However,  the order of the integral time scale is the same for both components. Without discussing the various estimates that can be evaluated using different criteria, 
we identify the two integral time scales around a value $\tau_L \simeq 5$ days, with a difference between zonal and meridional components of order $O(1)$ days. This implies that, for time lags significantly larger than $\tau_L$, the Lagrangian velocities along a trajectory become practically uncorrelated.      

\begin{figure}[!t]
\begin{center}
\includegraphics[clip=true,width=0.7\textwidth,angle=0]{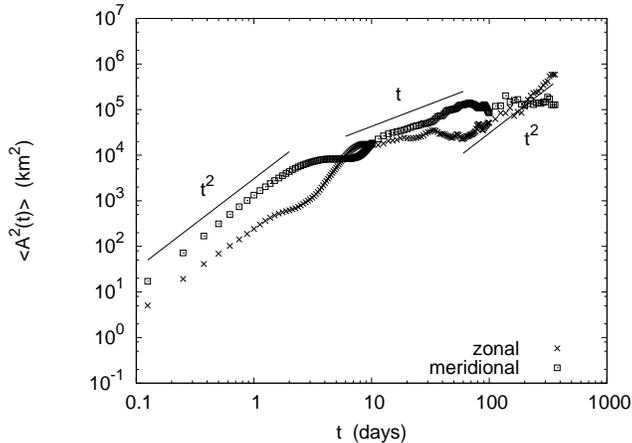}
\end{center}
\caption{Mean square drifter displacement from the mass center position, $\langle A^2(t) \rangle$, in zonal and meridional components.
The time sampling is $\Delta t=1/8$~day.}
\label{fig:absdisp}
\end{figure}

Absolute dispersion, i.e. the mean square fluctuation around the mean position, for zonal and meridional coordinates is reported in Fig. \ref{fig:absdisp}.  

In the present work we are not interested in reproducing estimates of small scale diffusivity, obtained by reconstructing 
a mean velocity field, in general not spatially uniform, and calculating the turbulent components from the difference between the drifter local velocities and the local mean field. Wherever we speak of diffusion coefficients, referring to  (\ref{eq:absdisp}) or (\ref{eq:reldisp}), we mean the {\it effective} diffusion coefficients of order 
$\sim O(UL)$ where $L$ and $U$ are determined, respectively, by the characteristic size and rotation velocity of the largest eddies. 
 Absolute dispersion, being dominated by the large scale features of the velocity field, 
reflects the anisotropy of the currents encountered by the drifters. The ballistic dispersion $\sim t^2$ 
scaling is plotted as reference for autocorrelated motions on time lags smaller than 10 days. The diffusion 
$\sim t$ scaling is a good approximation for meridional dispersion on time lags within $10-60$ days, but it is not good  
for zonal dispersion on any time lag. After about $60$ days, meridional dispersion tends to saturate, while zonal dispersion starts growing again as $\sim t^2$, 
likely because, on those time lags, on average, the drifters have reached the BMC, have stopped their southward meridional run along the BC and start moving eastward along the SAC, the natural consequence of which is the saturation of the meridional component and a second ballistic (or shear) regime for the zonal component. 

\subsection{Two-particle statistics}
\label{sec:2part}

\begin{figure}[!t]
\begin{center}
\includegraphics[clip=true,width=0.7\textwidth,angle=0]{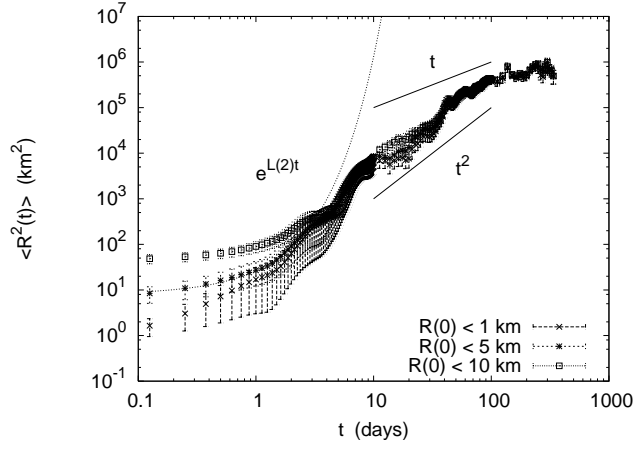}
\end{center}
\caption{Drifter relative dispersion $\langle R^2(t) \rangle$ for initial separations $R(0) \leq 1$ km, $R(0) \leq 5$ km and 
$R(0) \leq 10$ km. The quantity $L(2) \simeq 1.2$~day$^{-1}$ corresponds to $\lambda_L \simeq 0.6$~day$^{-1}$.  
The time sampling is $\Delta t=1/8$~day. Error bars are the standard deviations on the mean values.}
\label{fig:reldisp}
\end{figure}

Relative dispersion for three different initial separations is shown in Fig. \ref{fig:reldisp}.  Distances between two points on the ocean surface are calculated 
as great circle arcs, according to the spherical geometry standard approximation.  
\begin{figure}[!b]
\begin{center}
\includegraphics[clip=true,width=0.7\textwidth,angle=0]{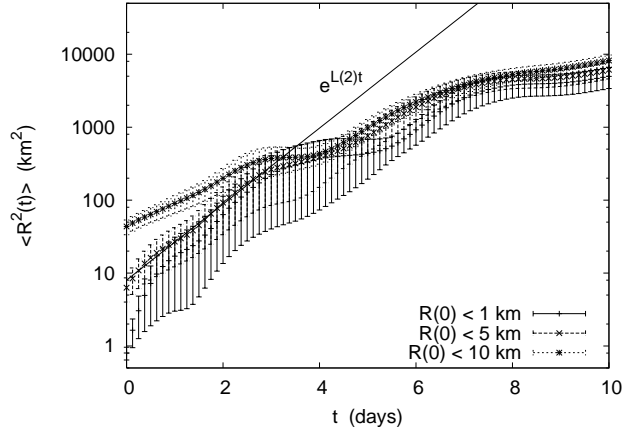}
\end{center}
\caption{Early regime of drifter relative dispersion $\langle R^2(t) \rangle$, for initial separations $R(0) \leq 1$, $\leq 5$ and 
$R(0) \leq 10$ km, with $L(2) \simeq 2 \lambda_L \simeq 1.2$~day$^{-1}$. The time sampling 
is $\Delta t=1/8$~day. Error bars are the standard deviations on the mean values.}
\label{fig:lyapexp}
\end{figure}
A way to increase relative dispersion statistics is to measure (\ref{eq:reldisp}) also for so-called chance pairs, i.e., 
particle pairs which may be initially very distant from each other but that, following the flow, happen to get sufficiently close to each other at
some later time after their release, in some random point of the domain (LaCasce 2008).  
The maximum number of pairs counted in the statistics depends on the initial threshold: 24 pairs for $R(0) \leq 1$ km, 30 pairs for $R(0) \leq 5$ km and 
39 pairs for $R(0) \leq 10$ km. 
The dependence of $\langle R^2(t) \rangle$ on $R(0)$ is well 
evident.  The early regime of relative dispersion is reported in Fig. \ref{fig:lyapexp}. The value of the MLE 
$\lambda_L \simeq 0.6$~day$^{-1}$  is of the same order as other estimates obtained in various ocean regions, as discussed later. 

The exponential growth has been adapted to the $R(0) \leq 5$ km curve which has a fair compromise between having 
a sufficiently small initial separation and, at the same time, an acceptable pair statistics. 

\begin{figure}[!t]
\begin{center}
\includegraphics[clip=true,width=0.7\textwidth,angle=0]{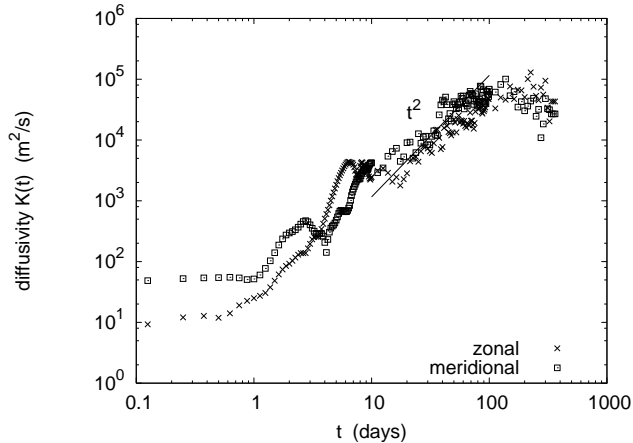}
\end{center}
\caption{Drifter relative diffusivity $K(t)$ in zonal and meridional components, for initial separations $R(0) \leq 5$ km. The $t^2$ scaling 
is plotted as reference to the Richardson regime. The time sampling is $\Delta t=1/8$~day.}
\label{fig:timediff}
\end{figure}

Relative diffusivity in the zonal and meridional directions, for initial separations $R(0) \leq 5$ km, is plotted in Fig. \ref{fig:timediff}. 
Although the curves do not display a clean scaling behavior, in the time range $10-100$ days the diffusivities show some resemblance with the 
$t^2$ law expected in the Richardson dispersion regime. 
We will examine now the results obtained with scale-dependent indicators. 

\begin{figure}[!t]
\begin{center}
\includegraphics[clip=true,width=0.7\textwidth,angle=0]{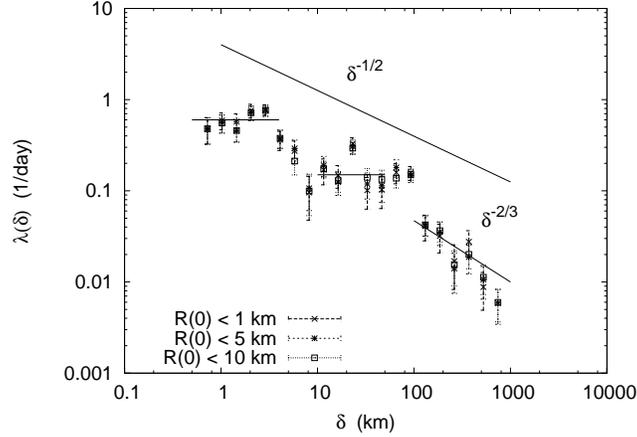}
\end{center}
\caption{FSLE for $R(0) \leq 1$, $\leq 5$ and $\leq 10$ km and amplification factor $\rho=\sqrt{2}$. The small scale 
plateau corresponds to exponential separation with rate $\lambda_L \simeq 0.6$~day$^{-1}$. 
The $\sim \delta^{-2/3}$ scaling corresponds to the Richardson law. The $\delta^{-1/2}$ scaling corresponds to a $k^{-2}$ spectrum. Error 
bars are the standard deviations on the mean values.}
\label{fig:fsle}
\end{figure}

The FSLE has been evaluated for the same initial thresholds 1, 5 and 10 km; the results are reported in Fig. \ref{fig:fsle}.     
Particle pairs are counted in the statistics if, at a given time, they happen to 
to be separated by a distance smaller than the fixed threshold.
The density of scales $\delta$ is fixed by setting $\rho=\sqrt{2}$. 
The maximum number of pairs considered varies with the initial threshold:  24 pairs for  $R(0) \leq 1$ km, 30 pairs for $R(0) \leq 5$ km and 
39 pairs for $R(0) \leq 10$ km. 
Larger initial thresholds (viz., $\geq 100$ km) could increase the statistics of an order $\sim O(10)$, but 
we will not consider this opportunity since, with too large initial thresholds, mesoscale dispersion rates may be affected by the inhomogeneity 
of the flow. In other words, at a certain separation scale 
two drifters might be still spatially correlated in some regions of the domain, but totally uncorrelated
in some other ones, depending on the structure of the flow.    
The parameter $\rho$ cannot be chosen arbitrarily close to 1 for obvious limitations because of finite resolution of the data. We have checked 
that, with $\rho=\sqrt{2}$, ``clipping events'', i.e., suspect transitions between two neighbouring scales in only one time step $\Delta t=1/8$~days, are not significant.   
Taking a value $\delta_R \simeq 30$~km for the Rossby radius, we can see that mesoscale dispersion (scales $> \delta_R$) is compatible with a 
$\delta^{-2/3}$ Richardson regime, as it occurs in the inverse cascade range of QG turbulence. Cut-off due to finite lifetime of the trajectories occurs 
before reaching $\sim 10^3$~km separations. The range $(1-100)$ km is characterized by a step-like behavior of the FSLE. The first plateau in the 
submesoscale range $\sim O(1)$~km corresponds to exponential separation with $\lambda \simeq 0.6$~day$^{-1}$, in agreement with the result obtained 
before for $\langle R^2(t) \rangle$.  
The second plateau at $\simeq 0.15$~day$^{-1}$  on $\sim O(\delta_R)$ scales is compatible with exponential separation due to a $k^{-3}$ direct cascade of QG turbulence. 
It is unclear why there exist two quasi-constant levels for $\lambda(\delta)$,  
which produce a significant variation of the exponential growth rate 
when $\delta$ decreases from meso to submesoscales.  Possible interpretations of this behavior may be found in the action of submesoscale vortices, even though the FSLE 
does not display a continuous cascade connecting these two scale ranges. The $\sim \delta^{-1/2}$ scaling is plotted as a reference to a $k^{-2}$ energy spectrum.  
Another remark concerns the gap at $\delta \sim 100$~km between the end of the plateau and the beginning of the $\sim \delta^{-2/3}$ regime. Inverse cascade implies 
that smaller eddies, formed by instability at the Rossby scale, tend to merge together in larger eddies up to scales in the range $(10^2-10^3)$ km.
Assuming this actually occurs, we also have to consider that the BC and BMC generate rings of size $\sim O(10^2)$ km which detach from 
the main currents. Such mesoscale rings can have trapping effects on the drifters, at least to some extent, which would determine an abrupt fall of the dispersion rate (FSLE).  The presence of the BMC can also have a role in temporarily inhibiting the growth of the two particle separation, until the drifters find their way 
eastward along the SAC.         

\begin{figure}[!t]
\begin{center}
\includegraphics[clip=true,width=0.7\textwidth,angle=0]{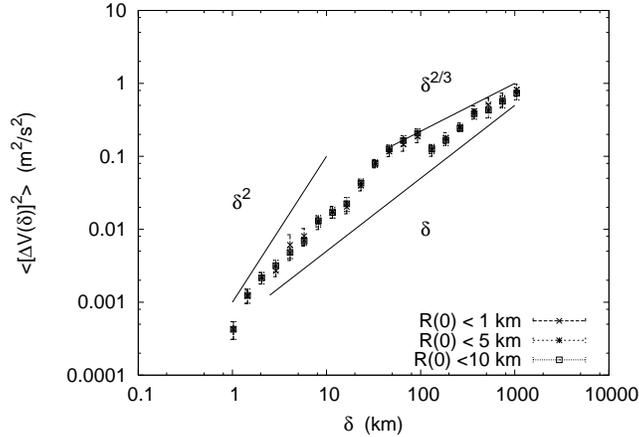}
\end{center}
\caption{FSRV computed together with the FSLE.  The FSRV is the mean exit velocity of the distance between two drifters from the ``shell'' $\delta$. The scalings 
$\delta^2$, $\delta^{2/3}$ and $\delta$ correspond to $k^{-3}$, $k^{-5/3}$ and $k^{-2}$ energy spectra, respectively. Error bars are the standard deviations 
on the mean values. 
}
\label{fig:fsrv}
\end{figure}

\begin{figure}[!t]
\begin{center}
\includegraphics[clip=true,width=0.7\textwidth,angle=0]{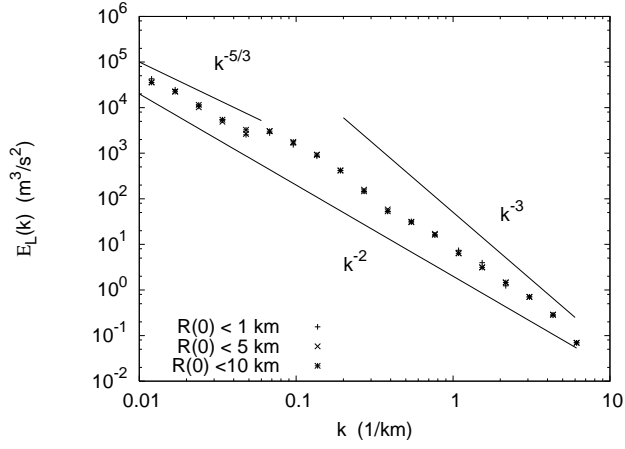}
\end{center}
\caption{Equivalent Lagrangian spectrum defined from the FSRV. The inverse cascade ($k^{-5/3}$), direct cascade ($k^{-3}$) and the submesoscale ($k^{-2}$) 
spectra are plotted as reference. The Rossby radius $\simeq$ 30 km corresponds to a wavenumber $k \simeq 0.2$. 
}
\label{fig:spectrum}
\end{figure}

Let us now consider the finite-scale (mean square) velocity difference, Fig. \ref{fig:fsrv}. The Lagrangian velocity components are reconstructed from the 
trajectories by means of a finite difference scheme.  
We notice, first, that the FSRV returns back a slightly cleaner picture than other indicators. This may depend, for instance, 
on the fact that the amplification time of the separation between two drifters passing from $\delta$ to $\rho \cdot \delta$ is affected by large fluctuations due to the 
non homogeneous characteristics of the flow, while the (mean square) relative velocity at scale $\delta$ is less affected by the history of the dispersion between two scales. The two ``valleys'' at nearly 10 and 100~km leave room to hypothesis of trapping events by structures of size comparable with those scales. 
The FSRV is consistent with Richardson dispersion on scales larger than the Rossby radius (except for the ``valley'' at 100~km);  
with exponential separation for two subranges $\sim (10-50)$ km and $\sim (1-5)$ km; with a $\delta$ scaling in the subrange $\sim (5-10)$~km, in turn 
consistent  with a $k^{-2}$ energy spectrum. The $\sim (10-50)$ km exponential regime can be associated to a $k^{-3}$ direct cascade, while the $\sim (1-5)$ km exponential 
regime can be associated to the action of submesoscale vortices.  From the FSRV an ``equivalent Lagrangian spectrum'' 
$E_L(k)=\langle \left[ \Delta V(k) \right]^2 \rangle / k$ can be defined, by dimensional arguments, replacing 
$\delta$ with $2\pi/k$, Fig. \ref{fig:spectrum}. 
The same scenario formerly indicated by the FSRV is reproduced, in $k$ space, by $E_L(k)$ as well.   

\begin{figure}[!t]
\begin{center}
\includegraphics[clip=true,width=0.7\textwidth,angle=0]{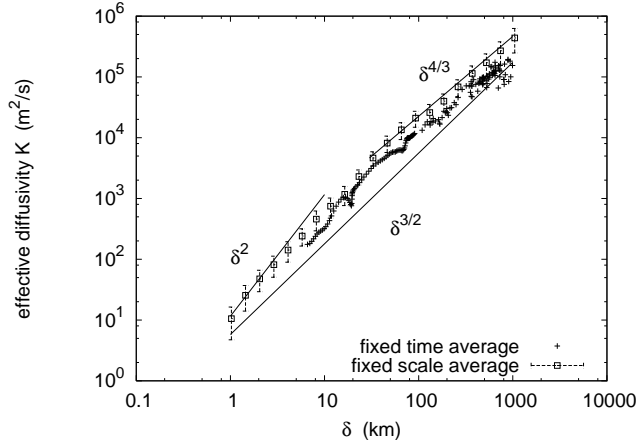}
\end{center}
\caption{Diffusivity as a function of the separation: fixed-time average $K(t)$ vs $\delta=\langle R^2(t) \rangle^{1/2}$ and fixed-scale average $K(\delta)$ vs $\delta$.   
The $\delta^{4/3}$ and $\delta^{2}$ correspond to $k^{-5/3}$ and $k^{-3}$ spectra, respectively. The $\delta^{3/2}$ scaling corresponds 
to a $k^{-2}$ spectrum.}
\label{fig:diff}
\end{figure}

We compare now the diffusivity (see Fig. \ref{fig:diff}) computed in both ways: as a fixed-time average (\ref{eq:diffusivity}) and as a fixed-scale 
average from the FSRV. 
Both quantities are plotted as functions of the separation between two drifters: $K(\delta)=(1/2) \cdot \delta \cdot \langle \left[ \Delta V(\delta) \right]^2 \rangle^{1/2}$ with 
$\delta$ as the independent variable, and $K(t)$ versus $\delta=\langle R^2(t) \rangle^{1/2}$ where the independent variable is the time $t$. 
The diffusivity is less affected by distortions present in the signal than other indicators. The two regimes $\sim \delta^{4/3}$ at mesoscales and 
$\sim \delta^{2}$ at scales $\sim O(\delta_R)$ are consistent with the inverse cascade $k^{-5/3}$ and direct cascade $k^{-3}$, respectively, as predicted by QG turbulence theory. 
A scaling $\sim \delta^{3/2}$, corresponding to a $k^{-2}$ spectrum, connects scales of the order of the Rossby radius to the submesoscale, and 
a scaling $\sim \delta^{2}$ is the signature of exponential separation driven by velocity field features on scales $\sim O(1)$ km.      

\section{Numerical simulations}
\label{sec:model}
To check the consistency of the results obtained from the data analyis of the MONDO drifters, we have performed two numerical experiments in which  
$\sim O(10^2)$ virtual drifters are "deployed" in a surface current field generated by a global operational forecast system using the Hybrid Coordinates Ocean Model (HYCOM; Bleck and Benjamin 1993) and the Navy Coupled Ocean Data Assimilation (NCODA; Cummings 2005). The model grid step is $1/12^{\circ}$  (approximately $7$~km) and outputs are available with a 1 day time step. From the 32 layer global grid, the surface layer is extracted in the area within $45^{\circ}$-$20^{\circ}S$ and $60^{\circ}$-$30^{\circ}W$ for the same period of drifters trajectories (20 September 2007 - 21 October 2008). 
Conversion from Earth coordinates to meters is accomplished following the great circle distance approximation. The integration algorithm of the Lagrangian trajectories uses a fixed timestep $\Delta t=1/24$~day and tri-linear interpolations. 
The maximum integration time is  $400$~days but particles reaching the shoreline are stopped and removed from the subsequent integration.

We indicate the numerical experiments with E1 and E2. In the first one (E1) the drifters are 
uniformly deployed in an area of about $(10\times10)$~km$^2$ centered around a position 
corresponding to the mean initial location of MONDO drifters. The average initial distance between particle pairs is 
$\langle R(0) \rangle \simeq 5$~km. The lifetime of trajectories is between $150$~days and $200$~days. In the second experiment 
(E2) the initial distribution of the drifters is characterized by larger separations, namely comparable to the spacing of the numerical 
grid ($\sim 10$~km); the average initial distance between particle pairs is $\langle R(0) \rangle \simeq 40$~km and the duration 
of trajectories is $250-400$~days.

\begin{figure}[!t]
\begin{center}
\includegraphics[clip=true,width=0.7\textwidth,angle=0]{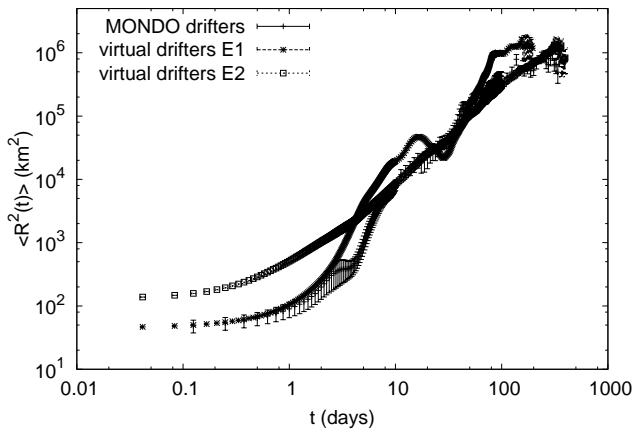}
\end{center}
\caption{Relative dispersion $\langle R^2(t) \rangle$ for initial separations $R(0) \leq 10$~km for MONDO drifters 
and virtual drifters from numerical experiments E1 and E2. For virtual drifters errors are of the order of point size; time sampling 
is $\Delta t=1/8$~day for MONDO drifters and $\Delta t=1/24$~day for virtual drifters.}
\label{fig:reldisp_vd}
\end{figure}

Here we present some comparisons between the analysis of data from model trajectories and those from the actual drifters, 
focusing on the relative dispersion process. First we consider a fixed-time indicator, namely relative dispersion $\langle R^2(t) \rangle$, 
for trajectories selected with small enough initial separation, that is for $R(0)<10$~km. As it can be seen in Fig.~\ref{fig:reldisp_vd}, 
the model outcome is in reasonable agreement with the behaviour of actual drifters. In particular, experiment E1 is capable of 
reproducing the early growth of relative dispersion but it cannot catch the late time behaviour due to too short trajectories. On the other 
hand, with experiment E2 we obtain a nicer agreement at late times, thanks to the longer lifetime of virtual trajectories in this case, 
but we loose the agreement at early times because now the initial separation between particle pairs is on average much larger.   

\begin{figure}[!t]
\begin{center}
\includegraphics[clip=true,width=0.7\textwidth,angle=0]{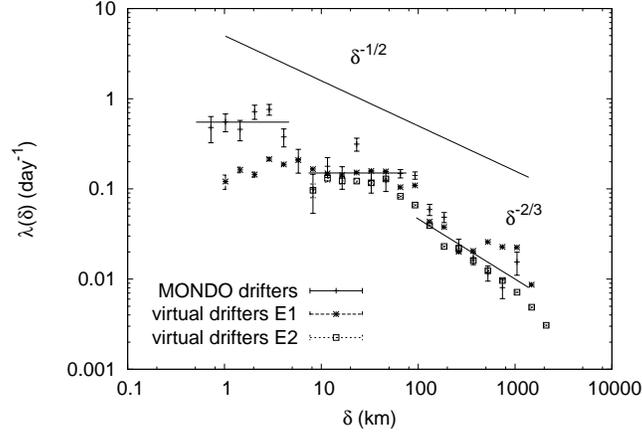}
\end{center}
\caption{FSLE for $R(0) \leq 50$~km and amplification factor $\rho=\sqrt{2}$ for MONDO drifters 
and virtual drifters from numerical experiments E1 and E2. For virtual drifters errors are of the order of point size. 
The large-scale saturation (E1) depends on the value of the trajectory integration time.}
\label{fig:fsle_vd}
\end{figure}

\begin{figure}[!b]
\begin{center}
\includegraphics[clip=true,width=0.7\textwidth,angle=0]{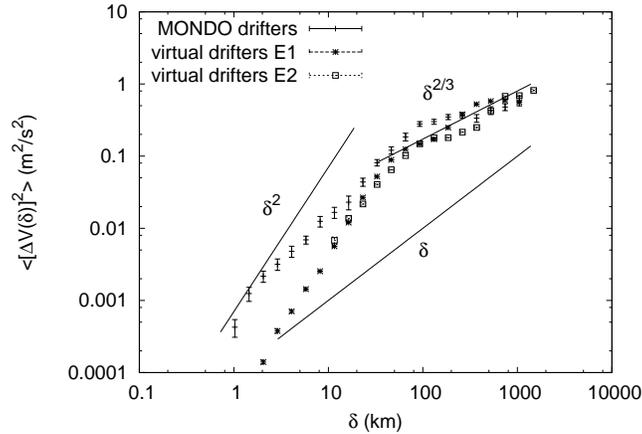}
\end{center}
\caption{FSRV for $R(0) \leq 50$~km for MONDO drifters and virtual drifters from numerical experiments E1 and E2. 
For virtual drifters errors are of the order of point size.}
\label{fig:fsrv_vd}
\end{figure}

For what concerns fixed-scale indicators, we consider the FSLE and the FSRV (Figures~\ref{fig:fsle_vd}, \ref{fig:fsrv_vd}).  
To increase the statistics we now select trajectories with a larger initial separation, that is for $R(0)<50$~km 
(similar results are found for smaller values of $R(0)$, though they are more noisy). The behaviours of both FSLE and FSRV support 
a double cascade scenario on the same scales as those found with MONDO drifters. The plateau value of FSLE at scales 
$O(\delta_R)$ is very close to the one found in the real experiment ($\lambda(\delta) \simeq 0.15$~day$^{-1}$). At larger scales, 
for both numerical experiments E1, E2 the behaviour of FSLE is compatible with a scaling law $\lambda(\delta) \sim \delta^{-2/3}$ 
supporting an inverse energy cascade process. Experiment E2, which is characterized by longer trajectories, shows 
a clearer scaling, thanks to a larger number of pairs reaching this range of large scales. Mean square velocity differences 
in the same range of scales display a reasonably clear $\delta^{2/3}$ scaling, also supporting an inverse energy cascade, 
with values close to those found with MONDO drifters. At separations smaller than the Rossby deformation radius, both indicators 
point to the presence of a direct enstrophy cascade: the FSLE is constant and the FSRV behaves as 
$\langle [\Delta V(\delta) ]^2\rangle \sim \delta^2$. This only partially agrees with the results found for real drifters, namely only 
in the scale range $10$~km~$<\delta<50$~km. At subgrid scales, velocity field features are not resolved and relative 
dispersion is necessarily a nonlocal exponential process driven by structures of size of the order of (at least) the Rossby radius. 
Correspondingly, the FSLE computed on model trajectories does not 
display the higher plateau level at scales smaller than $10$~km. 

\begin{figure}[htbp]
\begin{center}
\includegraphics[clip=true,width=0.7\textwidth,angle=0]{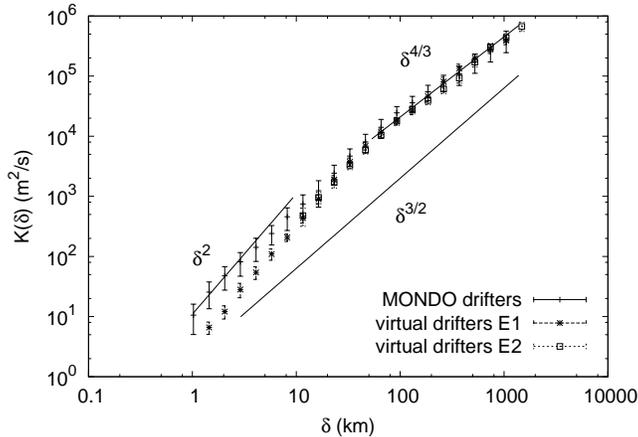}
\end{center}
\caption{Diffusivity $K(\delta)$ as function of the separation $\delta$ for MONDO drifters and virtual drifters from numerical 
experiments E1 and E2; here $R(0)<50$~km. }
\label{fig:diffu_vd}
\end{figure}

Finally, the behaviour of the relative diffusivity $K(\delta)$ as a function of the separation $\delta$ is shown in 
Fig.~\ref{fig:diffu_vd} for both MONDO and virtual drifters. Here $K(\delta)$ is computed from the mean square velocity 
difference, as described in Sections~\ref{sec:diffusion}c and~\ref{sec:results}b. Model and experimental data again are 
in agreement at scales larger than the numerical space resolution ($\sim 10$~km). Indeed, in both numerical experiments 
E1 and E2 we find scaling behaviours compatible with a QG double cascade: $K(\delta)\sim\delta^{4/3}$ (corresponding 
to Richardson's super-diffusion in an inverse cascade regime) for $\delta>\delta_R$, and $K(\delta)\sim \delta^2$ 
(corresponding to a direct cascade smooth flow) for $\delta<\delta_R$. At variance with the outcome of the real experiment 
with MONDO drifters, here we are unable to detect any significant deviations from the QG turbulence scenario at small scales.

The failure of the model to reproduce flow features at very small scale is of course due to its finite spatial 
resolution, which is of the order of $10$~km. Below this length scale, the velocity field computed in the model is smooth, while 
the one measured in the real ocean clearly displays active scales also in the range $(1-10)$~km. Nevertheless, the overall 
conclusion we can draw from the above comparisons is that the characteristics of the relative dispersion process found with MONDO 
drifters agree with those obtained with an Ocean General Circulation Model (OGCM) for scales $\delta>10$~km. 
\section{Discussion and conclusions}
\label{sec:conclusions}

Lagrangian dispersion properties of drifters launched in the southwestern corner of the South Atlantic Subtropical Gyre 
have been analyzed through the computation of  time-dependent and scale-dependent indicators.  
  
The data come from a set of 37 WOCE SVP drifters deployed in the Brazil Current around ($24^{\circ}$S, $44^{\circ}$W) during Monitoring by Ocean Drifters (MONDO) Project, an oceanographic campaign planned by PROOCEANO and supported by ENI Oil do Brasil in relationship with an oil drilling operation. 
The experimental strategy of deploying part of the drifters in 5-element clusters, with initial separation between drifters smaller than 1 km, allows to study 
relative dispersion on a wide range of scales. 

Single-particle analysis has been performed by computing classic quantities like Lagrangian autocorrelation functions and absolute dispersion, defined as the 
variance around the drifter mean position, as a function of the time lag from the release. Velocity variances ($\sim 500$~cm$^2$~s$^{-2}$) and integral time scales 
($\tau_L \sim 5$~days) are compatible with the estimates obtained in the analysis of  the FGGE drifters 
(Figueroa and Olson 1989; Sch\"afer and Krauss 1995). 
Anisotropy of the flow is reflected in the different behavior of the zonal and meridional components of the absolute dispersion. Being the MONDO drifters 
advected mostly by the boundary currents surrounding the Subtropical Gyre, the Brazil Current first and the South Atlantic Current later, the meridional component of the absolute dispersion is dominant as long as the mean drifter direction is nearly southward (BC), while the zonal component dominates (see the appearance of the late ballistic regime) as the  mean drifter direction is nearly eastward (SAC). Early time advection is modulated, also, by the response of the 
currents to a wind forcing of period $\sim 6.5$~days, a characteristic meteorological feature of the BC dynamics.  

Two-particle analysis has been performed by means of both fixed-time and fixed-scale averaged quantities. Classic indicators like the mean square relative 
displacement and the relative diffusivities between two drifters as functions of the time lag from the release give loose information about early phase 
exponential separation, characterized by a mean rate $\lambda_L \simeq 0.6$~day$^{-1}$, and long time dispersion approximated, to some extent, by Richardson 
super-diffusion before the cut-off due to the finite lifetime of the trajectories. Evidence of a small scale exponential regime for relative dispersion is common 
to other drifter studies for different ocean regions (LaCasce and Ohlmann 2003; Ollitrault et al. 2005, Koszalka et al. 2009, 
Lumpkin and Elipot 2010).    
Scale-dependent indicators return back a cleaner picture, compatibly with the limited statistics allowed by the experimental data and the non homogeneous and non 
stationary characteristics of the flow.   

The FSLE displays a mesoscale [$(100-500)$~km] regime compatible with Richardson super-diffusion, Lagrangian counterpart of the 
2D inverse cascade scenario characterized by a $k^{-5/3}$ energy spectrum.  At scales smaller than 100 km the FSLE has a step-like shape, with a first plateau 
at level $\sim 0.6$~day$^{-1}$ in the submesoscale range $\sim O(1)$ km, and a second plateau at level $\simeq 0.15$~day$^{-1}$ for scales comparable 
with the Rossby radius of deformation ($\delta_R \simeq 30$ km). Constant FSLE in a range of scales corresponds to exponential separation. The $\sim O(\delta_R)$ plateau could be 
related to the 2D direct cascade characterized by a $k^{-3}$ energy spectrum, while the origin of the $\sim O(1)$ km plateau is likely related to the existence of 
submesoscale features of the velocity field, the role of which has been recently assessed by means of high-resolution 3D simulations of upper ocean turbulence at 
Rossby numbers $R_o \sim O(1)$  (Capet et al. 2008a,b; Klein et al. 2008). The FSLE does not display a clean continuous cascade scaling $\sim \delta^{-1/2}$, 
corresponding to a $k^{-2}$ energy spectrum, from sub to mesoscales; however, it highlights the existence of scales of motion hardly reconcilable 
with the QG turbulence scenario, as analogously assessed also by Lumpkin and Elipot (2010) for drifter dispersion in the North Atlantic. 
 
The FSRV measures the mean square velocity difference at scale $\delta$. The scaling of the FSRV is related to the turbulent 
characteristics of relative dispersion. The FSRV behavior is, under a certain point of view, cleaner than that of the FSLE, but it is affected by the presence of 
two "valleys", roughly at $\simeq 10$ km and at $\simeq 100$ km, likely associated to trapping events (the same features are present also in the FSLE 
behavior). Coherent structures on scales $\sim 10$ and $\sim 100$~km may be responsible of the ``fall'' in the relative dispersion rate. These scales 
are of the order of the Rossby radius and of the mesoscale rings that detach from the current systems, respectively. We must also consider the role of the 
Brazil-Malvinas Confluence which tends to inhibit the growth of the dispersion as the drifters, initially flowing southwestward along the Brazil Current, encounter 
the northeastward flowing Malvinas Current before being, eventually, transported eastward along the South Atlantic Current.    
The FSRV displays a mesoscale $\sim \delta^{2/3}$ scaling (except for the ``valley''),  compatible with a $k^{-5/3}$ inverse cascade; a $\sim \delta^2$ scaling 
in the two subranges $\sim (10-50)$~km and $\sim (1-5)$~km, which correspond to the step-like shape of the FSLE; a $\sim \delta$ scaling which, to some extent, 
is compatible with a $k^{-2}$ regime connecting scales of the order of the Rossby radius to the submesoscale (below which the velocity field becomes approximately 
smooth). The equivalent Lagrangian spectrum $E_L(k)$, formed by dividing the FSRV (essentially the relative kinetic energy at scale $\delta$)  by the wavenumber $k=2\pi/\delta$, returns back the same scenario in $k$ space.   

Last, we have compared the scale-dependent relative diffusivity $K(\delta)$ constructed from the FSRV, for which $\delta$ is the independent variable, 
with the classic relative diffusivity $K(t)$ seen as a function of the mean separation between two drifters $\langle R^2(t) \rangle^{1/2}$, for which 
$t$ is the independent variable. What emerges from the analysis of the diffusivities is that, in the mesoscale range, loosely from the Rossby radius 
up to scales $\sim O(10^2)$~km, $K(\delta) \sim \delta^{4/3}$, corresponding to the $k^{-5/3}$ inverse cascade; in the submesoscale range $\sim (1-5)$~km, 
and in a limited subrange from about the Rossby radius down to $\sim 10$~km, $K(\delta) \sim \delta^2$, corresponding to exponential separation; in the 
subrange $\sim (5-10)$~km, the scaling $K(\delta) \sim \delta^{3/2}$ is compatible with a $k^{-2}$ regime. The $k^{-2}$ spectrum 
is a characteristic of upper ocean turbulence when the Rossby number is order $\sim O(1)$, 
as recently assessed with high-resolution 3D model simulations (Capet et al. 2008a,b; Klein et al. 2008), which connects the mesoscale 
to the submesoscale $\sim O(1-10)$ km (McWilliams 1985). Below the submesoscale the velocity field is reasonably assumed to vary smoothly. 
A rough estimate of the Rossby number associated to the MONDO drifter dynamics gives a value (at least) $R_o = U/(Lf) \sim O(10^{-1})$, taking  
$U \sim 0.3$ m/s, $L \sim 30$ km and $f \sim 10^{-4}$ 1/s, which is, nonetheless, considerably larger than the typical Rossby number in open ocean. 
 Although the drifter data analysis does not show a clear evidence of a relative dispersion regime corresponding to the $k^{-2}$ spectrum, the presence 
of velocity field features of size comparable to submesoscale vortices is reflected, to some extent, by the behavior of the small scale relative dispersion process.          

Numerical simulations of the Lagrangian dynamics have been performed with an OGCM of the South Atlantic (Huntley et al. 2010).  
The results concerning the relative dispersion essentially agree with the data analysis of the MONDO drifters, within the limits of the available numerical resolution. 
In particular, two-particle statistical indicators such as the FSLE, the FSRV and the scale-dependent relative diffusivity $K(\delta)$, 
computed on the trajectories of virtual drifters, display the same behaviour found for MONDO drifters for scales larger than 
approximatley $10$~km, that is larger than the numerical grid spacing. Below this length scale, evidently, the model flow field is smooth, 
hence relevant departures from QG turbulence and the role of submesoscale structures cannot be assessed. Further investigations 
on the modeling of submesoscale processes would provide extremely useful in order to make a clearer picture of the small scale dynamics of 
the surface ocean circulation in the region.\\ 

FDS thanks Eni Oil do Brasil S.A. in the person of Ms. Tatiana Mafra for being the financial promoter of MONDO
Project and for making the experimental data available to the scientific community. 
SB acknowledges financial support from CNRS. We thank D. Iudicone and E. Zambianchi for useful discussions and suggestions. 
The authors are grateful, also, to three anonymous Reviewers who have helped to improve the substance and the form of this work with their critical remarks.  

\end{document}